\documentclass[12pt,a4paper]{article}
\usepackage{amsmath,amssymb,graphicx,psfig,endfloat}
\begin{document}
%\begin{frontmatter}
\title{Feedback processes in cellulose thermal decomposition. Implications for
fire-retarding strategies and treatments.}

\author{R. Ball\thanks{Rowena.Ball@anu.edu.au} \\
Department of Theoretical Physics,
Australian National University\\ Canberra ACT 0200 Australia
\and
A. C. McIntosh$^{\scriptsize\text a}$ \and 
J. Brindley$^{\scriptsize\text b}$\\
$^{\scriptsize\mathrm a}$Department of Fuel and Energy,
$^{\scriptsize\mathrm b}$Department of Mathematics \\
University of Leeds, Leeds LS2 9JT U.K.}
%\date{}
\maketitle
%\newpage

\begin{abstract}
A simple dynamical system that models the competitive thermokinetics 
and chemistry of cellulose 
decomposition is examined, with reference to evidence from experimental
studies indicating that char formation is a low  activation energy exothermal
process and volatilization is a high activation energy endothermal
process. The thermohydrolysis chemistry at the core of the primary
competition is described. Essentially, the
competition is between two nucleophiles, a molecule of water and an -OH group
on C$_6$ of an end glucosyl cation, to form {\em either} a reducing chain fragment 
with the propensity to undergo the bond-forming reactions that ultimately 
form char {\em or} a levoglucosan-end-fragment that depolymerizes to volatile
products. The results of this analysis suggest that promotion of
char formation under thermal stress can actually {\em increase} the production 
of flammable volatiles.
Thus we would like to convey an important safety message in this paper: 
in some situations where heat and mass transfer is restricted in 
cellulosic materials, such as 
furnishings, insulation, and stockpiles, the use 
of char-promoting treatments for 
fire retardation may have the effect of increasing the risk of
flaming combustion.  
\end{abstract}
%\begin{keyword}
%\end{keyword}
%\end{frontmatter}

\renewcommand{\baselinestretch}{1.2}\normalsize
\section{Introduction}
\markboth{Feedback processes in cellulose thermal decomposition \ldots}
{Feedback processes in cellulose thermal decomposition \ldots}\rightmark
Thermokinetic and chemical feedback 
processes that determine the course and outcome of cellulose thermal 
decomposition are of particular interest in the development of appropriate
flame-inhibiting treatments of manufactured goods  such
as bedding, furnishings, and insulation, or stockpiled raw cellulosic 
material such as bagasse or hay. The fuel that ignites in the flaming combustion of
cellulose substrates consists mostly of volatile, small-molecule substances that are 
supplied continuously by thermal degradation of the solid cellulose. In designing more 
efficient and effective fire-retardants for cellulose --- and, more generally, in
the pre-emptive control of cellulose combustion --- it is thus important to 
understand both the dynamics and the chemistry of decomposition.
  In this work we investigate
a phenomenonological model for the thermal 
decomposition of cellulose substrates that is based on the following 
two key aspects of
the known chemistry and thermokinetics: (1) involvement of water in promoting
char formation, and (2) reaction enthalpy feedback to promote 
volatilization. 

A great many thermogravimetry (TG) and differential scanning calorimetry 
(DSC) experiments, complemented by qualitative and quantitative product
analysis, have affirmed that cellulose 
thermal decomposition is largely a competitive process
\cite{Broido:1970,Arseneau:1971,Bilbao:1987a,Mok:1992,Williams:1996,Gupta:1999,Gronli:1999}. 
In the big picture that has taken shape from 
these results and from associated modelling 
studies (for example, see 
\cite{Bradbury:1979,Agrawal:1988a,Agrawal:1988b,Diebold:1994,Conesa:1995,Milo:1995} 
and the reviews in \cite{Antal:1995} and \cite{Diblasi:2000}) formation of volatiles and char
are seen to be reciprocally linked to some extent. 

Some details have also emerged 
of the chemistry involved in these 
competitive pathways \cite{Phillip:1984,Essig:1989,Price:1997}, 
although a complete picture of the complex reaction
network remains a major experimental challenge. 
What is clear though is that the involvement
of water in hydrolysis reactions is crucial in determining the rates and 
outcomes of the competitive pathways.
A thorough survey and compilation of the known chemistry and thermokinetics
of cellulose thermal decomposition was carried out  in 
\cite{Ball:1999b}, on the basis of which a dynamical model was
built up that includes the role of water in the reaction chemistry. 
In this paper we concentrate on the kinetic and thermal aspects of the
decomposition process,
emphasizing the role played by thermal feedback in regulating the
decomposition pathways. We demonstrate the potential for thermal instabilities
by considering the qualitative behaviour of a very simple thermokinetic model, 
which nevertheless contains the important rate processes in cellulose thermal
decomposition: competitive temperature-dependent
reaction rates, reaction enthalpy feedback, and heat losses.  
\section{\label{chemistry}Basic chemistry and phenomenology of cellulose decomposition}
\begin{figure}\hspace{1.5cm}
\includegraphics[width=0.8\hsize]{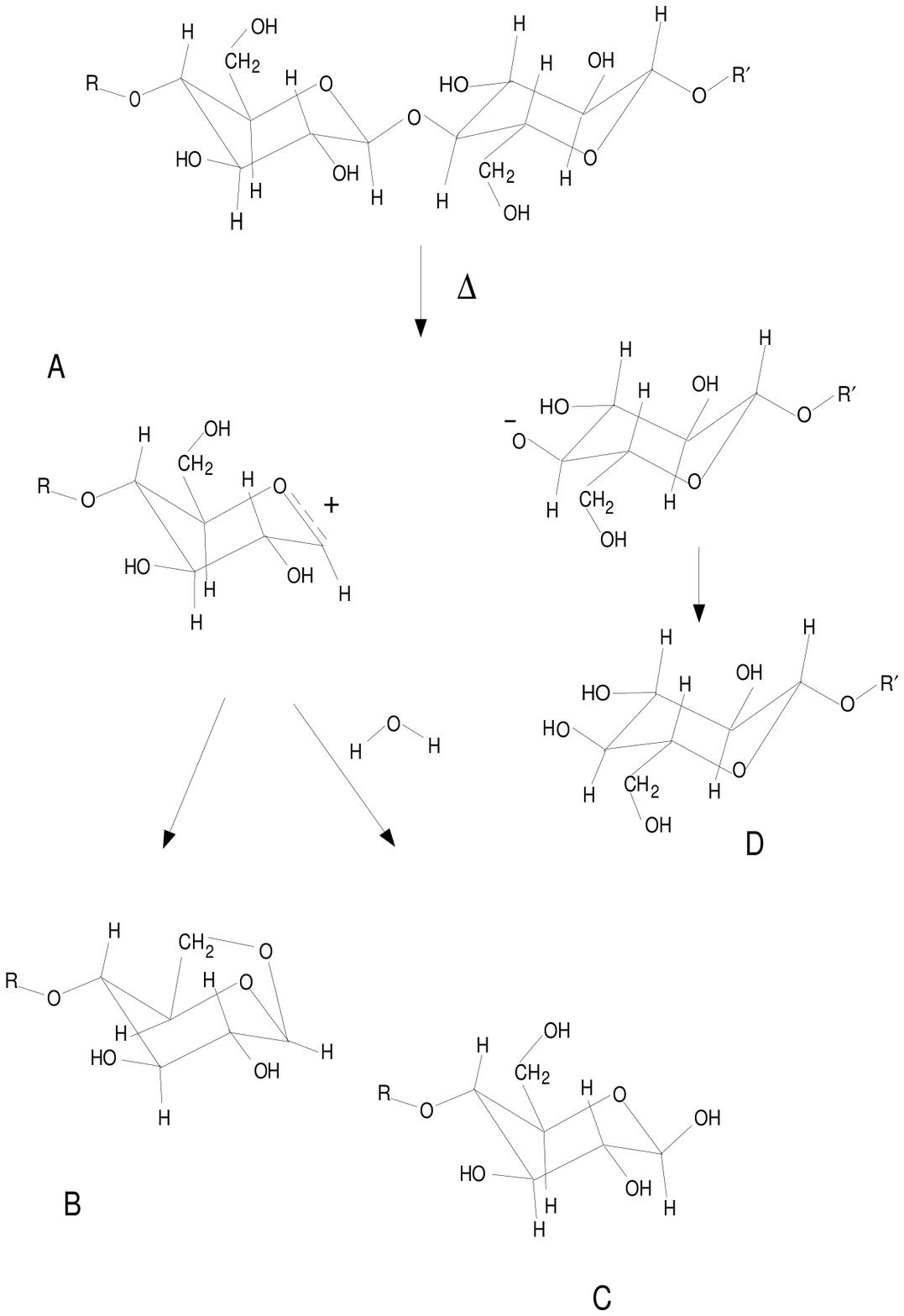}\vspace*{0.5cm}
\caption[The carbonium ion (A) may form a levoglucosan end (B) via 
intramolecular nucleophilic attack, or a reducing end (C) when intercepted by a 
water molecule. In both cases a non-reducing end (D) is also formed. ]
{\label{figure1}The carbonium ion (A) may form a levoglucosan end (B) via 
intramolecular nucleophilic attack, or a reducing end (C) when intercepted by a 
water molecule. In both cases a non-reducing end (D) is also formed.}
\end{figure}

The chemistry of the primary 
competitive reactions is illustrated in Fig~\ref{figure1}. 
The initial step is
believed to be heterolytic thermal scission of glycosidic linkages at random
chain locations in amorphous regions of the cellulose. In a dry environment 
the positively charged end (A in the figure) is rapidly cyclized to a 
levoglucosan end (B in the figure), with the -OH group on C$_6$ of the 
unit as nucleophile.
Thermolysis at the next glycosidic linkage of this fragment releases the
volatile levoglucosan. When water is present it can compete as a 
nucleophile for the positively charged centre on A to produce a reactive 
reducing end (C in the figure). It is this species which is believed to
undergo the subsequent dehydration, decarboxylation, and cross-linking
reactions that produce the char. (The negatively charged fragment of
the thermal scission rapidly picks up a positive hydrogen ion from the
-OH nucleophile group or from water to form a non-reducing end D. In general,
non-reducing ends of carbohydrates are relatively unreactive.) 

Empirical methods for producing good
yields of char by heating damp wood in an oxygen-deficient atmosphere
have been used for centuries. Systematic investigations of this effect
using DSC experiments were carried out by Mok {\it et al} \cite{Mok:1992}, 
who observed that
either a high concentration of vapour products or added water increased
char yield and decreased the temperature of onset of decomposition. 
Subsequent analysis of these data \cite{Varhegyi:1993} led to the  
suggestion that the water produced in thermal dehydration reactions
feeds back to hydrolyse the unreacted cellulose.
In \cite{Rubtsov:1993} and \cite{Szabo:1996} evidence was also presented from 
pyrolysis experiments that 
product water accelerated the formation of char. Some investigations of 
the role of the hydrolysis chemistry of cellulose under thermal stress in producing 
this effect were reported in \cite{Essig:1989} and \cite{Jakab:1997}. 

It is also known that covered stockpiles of cellulosic materials such
as bagasse, hay, and cotton undergo spontaneous combustion much more
readily when critically damp than in dry conditions. 
Dynamical models that reproduce qualitatively the behaviour that is
typical in damp combustion have been investigated by Gray \cite{Gray:1990}
and others \cite{Sisson:1992,Sisson:1993,McIntosh:1994b,McIntosh:1995,Chong:1999}. 
In these systems, which model the combustion subsequent to
decomposition, critical, unstable, or oscillatory thermal
behaviour can occur due to the coupling of two or more nonlinear 
temperature-dependent reaction rates with restricted linear heat losses. 
\section{Alternative pathways are thermally connected}
\begin{figure}
%\hspace*{2cm}\psfig{file=bmb2fig2.ps}\vspace*{1cm}
\hspace*{2cm}\includegraphics[width=0.7\hsize]{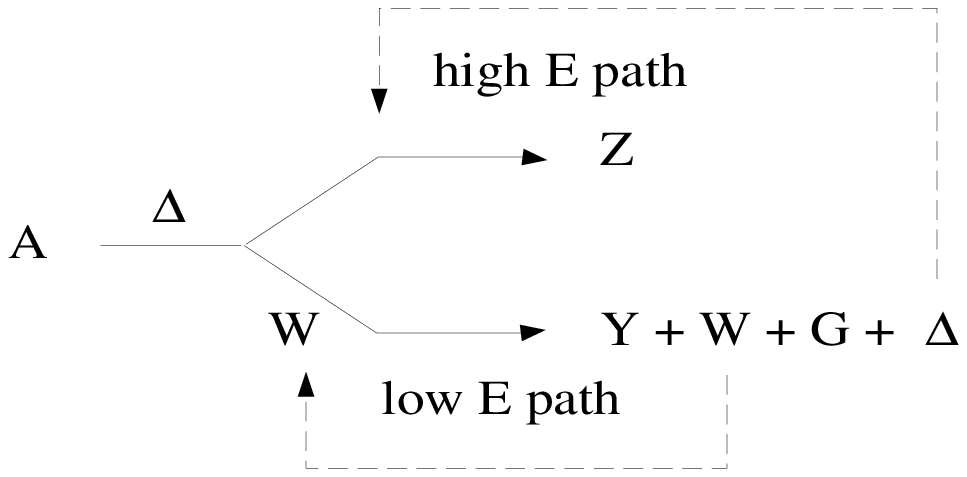}\vspace*{1cm}
\caption[How chemical and thermal feedback may be involved in
cellulose decomposition. A, Z, W, Y, G, and $\Delta$ stand for cellulose,
volatiles such as levoglucosan, water, dehydrated product or 
char, gases such as CO and CO$_2$, and heat, respectively]
{\label{figure2}How chemical and thermal feedback may be involved in
cellulose decomposition. A, Z, W, Y, G, and $\Delta$ stand for cellulose,
volatiles such as levoglucosan, water, dehydrated product or 
char, gases such as CO and CO$_2$, and heat, respectively.}
\end{figure}

The studies reviewed briefly in the previous section suggest the rather
counter-intuitive notion that  water can {\em increase} the
temperature of thermally decomposing cellulose. To guide our thinking on this,
we need to put together three pieces of information:\enlargethispage{5mm}
\begin{enumerate}
\item The volatilization 
rate is more temperature-sensitive, i.e., has higher activation energy,
than the charring rate. 
A typical experimental result that leads to this inference was reported in
\cite{Williams:1996}. It was found that as the pyrolysis temperature of 
pine wood, cellulose, hemicellulose, and lignin was increased, the 
percentage mass of solid char decreased, while that of 
gas and oil (volatile or tar) products increased. 
\item 
Estimates of the magnitude of the reaction enthalpies vary, but there
is agreement that the charring process is highly exothermal whereas 
volatilization is endothermal, at the very least to the extent of the enthalpy
of vapourization of levoglucosan 
(see for example 
\cite{Rubtsov:1993,Kung:1973,Mok:1983,Chen:1991,Milo:1996}). 
\item  As described in \cite{Ball:1999b}, water of dehydration can 
participate in hydrolysis that effectively blocks the cyclization to
levoglucosan.
\end{enumerate} 
Putting these together we can draw the schematic diagram 
of the competitive decomposition shown in Fig 
\ref{figure2}. 
As the substrate A is heated the low activation energy dehydration reactions 
are initiated. The water produced by this process can feed back into hydrolysis reactions
that inhibit volatilization and enhance char formation. However, the
charring reactions are bond-forming and heat-releasing. Under circumstances where there is no
efficient linear heat sink, the alternative,
high activation energy, volatilization reactions begin to take over. 
The heat requirement of volatilization means that it can self-damp, 
thus opening the reaction field again to the charring pathway. 
It is clear that there is potential for the system to attain a stable
steady state or limit cycle. In reality, of course, the situation is more complex --- and
probably more dangerous, because accumulation of hot volatiles in what is often a highly
oxidizing environment is likely to lead to spontaneous flaming ignition. 
 %(or other moisture in the system) 

\section{\label{proxy}A simple proxy for competitive thermal decomposition}
To strengthen our feel for the qualitative aspects of the  
system described above, we will study the thermokinetics of the 
following simple competitive reaction scheme:
\begin{align}
\textrm{A}&\quad\overset{R_w}{\longrightarrow}\quad \textrm{W}+\textrm{Y}\label{r1}\\
\textrm{A}+\textrm{W}&\quad\overset{k_1\left(T\right)}{\longrightarrow}
\quad \textrm{Y}+\Delta\label{r2}\\
\textrm{A}+\Delta&\quad\overset{k_2\left(T\right)}{\longrightarrow}
\quad \textrm{Z}\label{r3}.
\end{align}
This may be viewed as a model for the  
the thermal decomposition of cellulose, with
A, W, Y, Z, and $\Delta$ representing the substrate,  
water, dehydrated product or char, volatiles, and heat respectively. 
Although a gross simplification of the real process, it nevertheless
contains the important overall characteristics of temperature-dependent
reaction rates, reactive heat generation/consumption, and feedback of W.
Making the reasonable approximations that (1) the rate of supply $R_w$ of W
from reaction \ref{r1} is constant (the activation energy for this reaction is believed to
be very low) and thermally neutral, (2) 
A is drawn from an ``infinite pool'', and (3) the system is 
spatially homogeneous with linear heat dissipation, we can write the following  
dynamical equations to describe the evolution of 
the vapour-phase components and the temperature:
\begin{align}
\frac{dw}{d\tau}&=-e^{-1/u}w+r-fw\label{dwdt}\\
\frac{dz}{d\tau}&=\nu e^{-\mu/u}-fz\label{dzdt}\\
\bar{C}\frac{du}{d\tau}&=e^{-1/u}w-\alpha\nu e^{-\mu/u} +
\ell\left(u_a-u\right).\label{dudt}
\end{align}
The dimensionless groups and the quantities from which they are comprised 
are explained in the Appendix.  Briefly, $w$ is the concentration of reactant water, 
$z$ is the concentration of 
volatile product, $u$ is the dynamical 
temperature. Water is supplied by reaction \ref{r1} at a temperature-independent,
thermally neutral rate $r$, and is removed with the vapour-phase outflow at rate $f$ and 
by the hydrolysis process \ref{r2} that ultimately 
leads to char formation.  Volatiles are produced in process \ref{r3} and are also 
removed in the vapour-phase outflow. Process \ref{r2} is a heat source, process \ref{r3}
is a heat sink, and a linear heat exchange with rate coefficient $\ell$ is assumed to occur 
between the system and its immediate environment at constant ambient temperature $u_a$.

Eqs \ref{dwdt}--\ref{dudt} are based on a CSTR (continuous stirred tank reactor) paradigm
with two temperature-dependent reaction rates. Comparable dynamical models 
that include two or more temperature-dependent reaction rates
have been analysed in the literature from several points of view: 
fundamental studies of the dynamical and bifurcation behaviour 
may be found in \cite{Gray:1994} and \cite{Ball:1999}, applications to
the wet combustion problem were cited above, 
%(made in \cite{Chong:1999})
and their use in the 
the stabilization of thermal runaway reactions was described in
 \cite{Gray:1999}.
In all of these problems it was found that the coupling of two non-identical
Arrhenius terms could give rise to steady-state and oscillatory behaviour of 
greater complexity than in corresponding systems with a single Arrhenius term.
Concomitantly, it has been shown that an endothermal reaction rate 
(the second term on the right-hand side 
of Eq. \ref{dudt}) can stabilize the system temperature \cite{Gray:1999}. 
In this system the effects of the high activation energy endothermal reaction 
are modified by the rate $r$ at which water is fed into the exothermal 
reaction. 

\begin{figure}\hspace*{-1.5cm}
\hbox{\includegraphics[width=6cm]{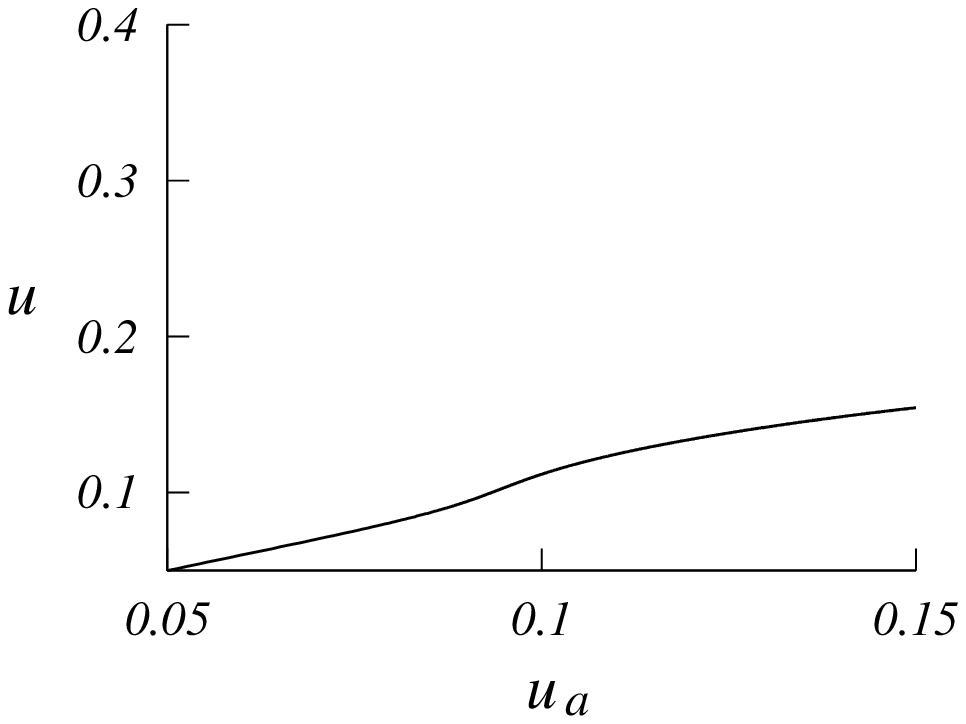}
     \hspace*{-0.5cm}\includegraphics[width=6cm]{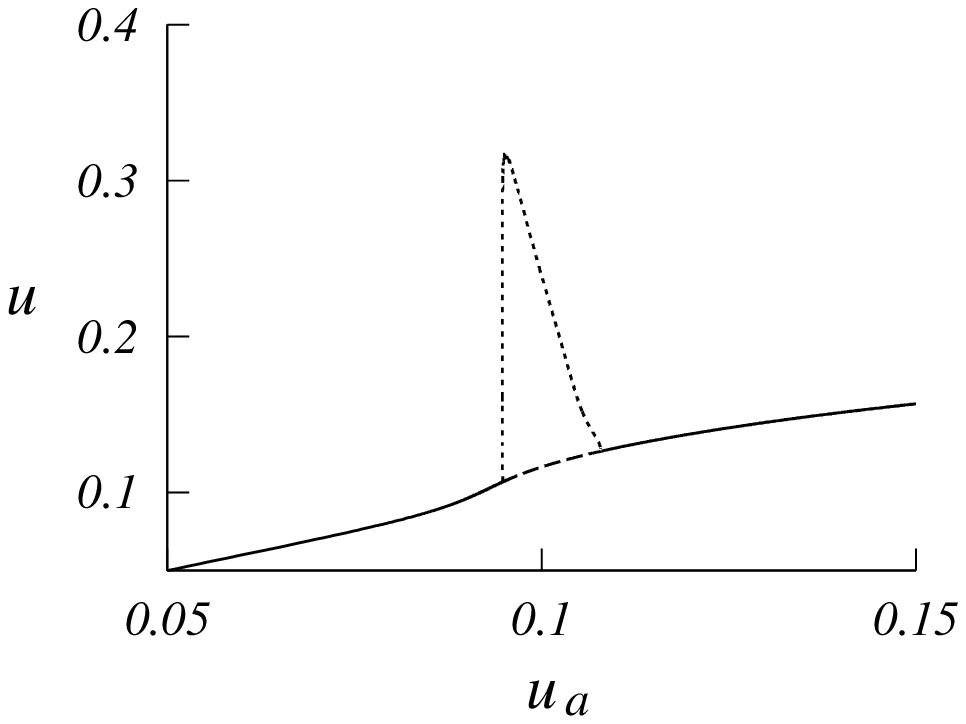}
      \hspace*{-0.5cm}\includegraphics[width=6cm]{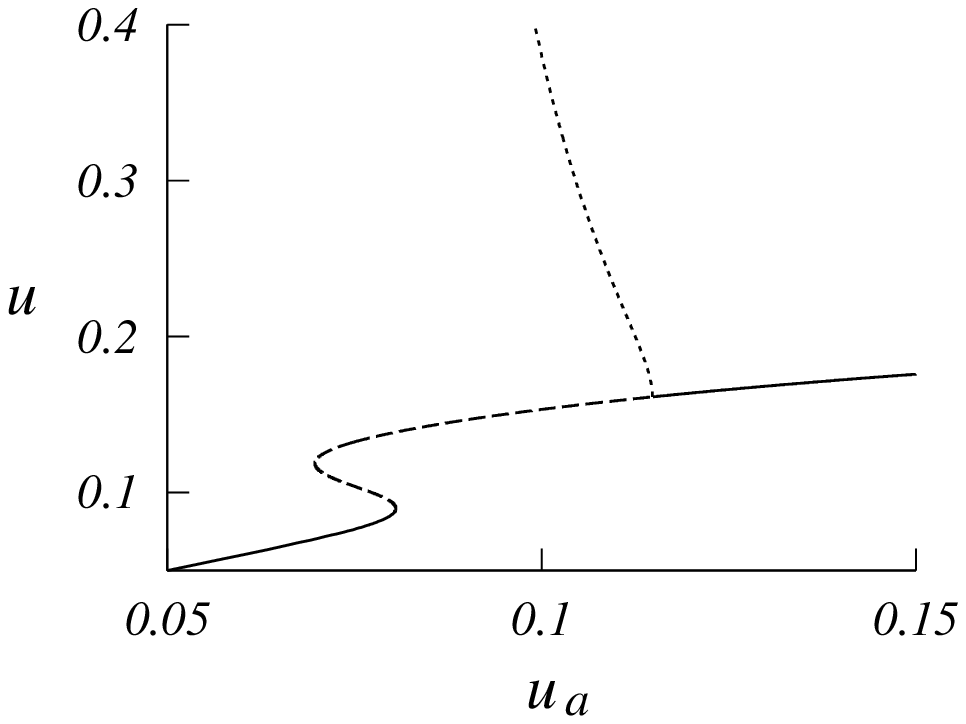}
}
       \hspace*{-1.5cm}
\hbox{\includegraphics[width=6cm]{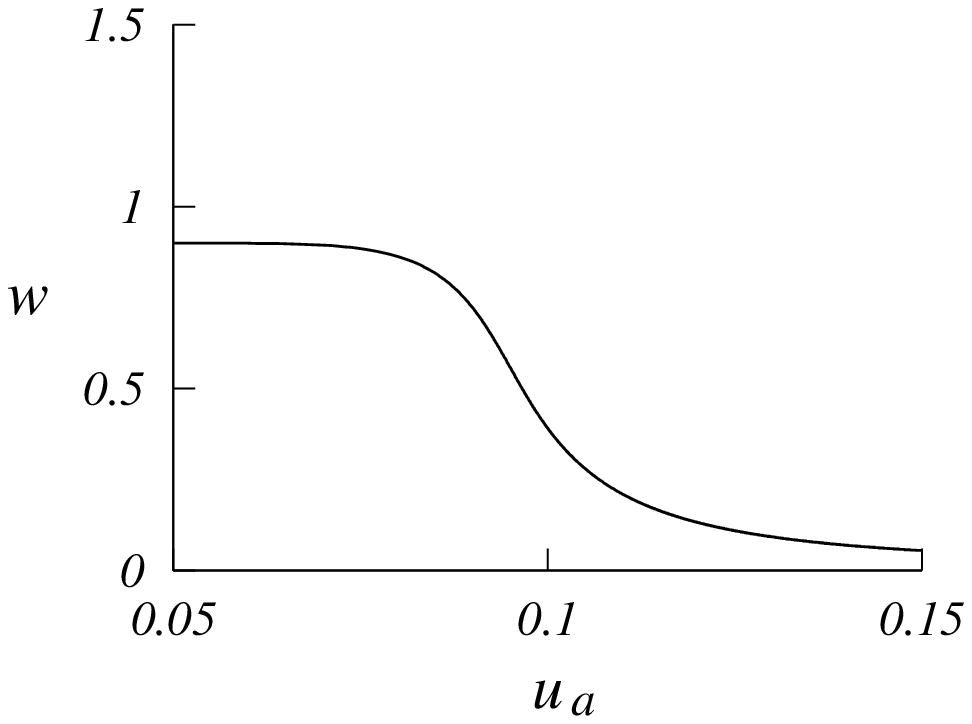}
      \hspace*{-0.5cm}\includegraphics[width=6cm]{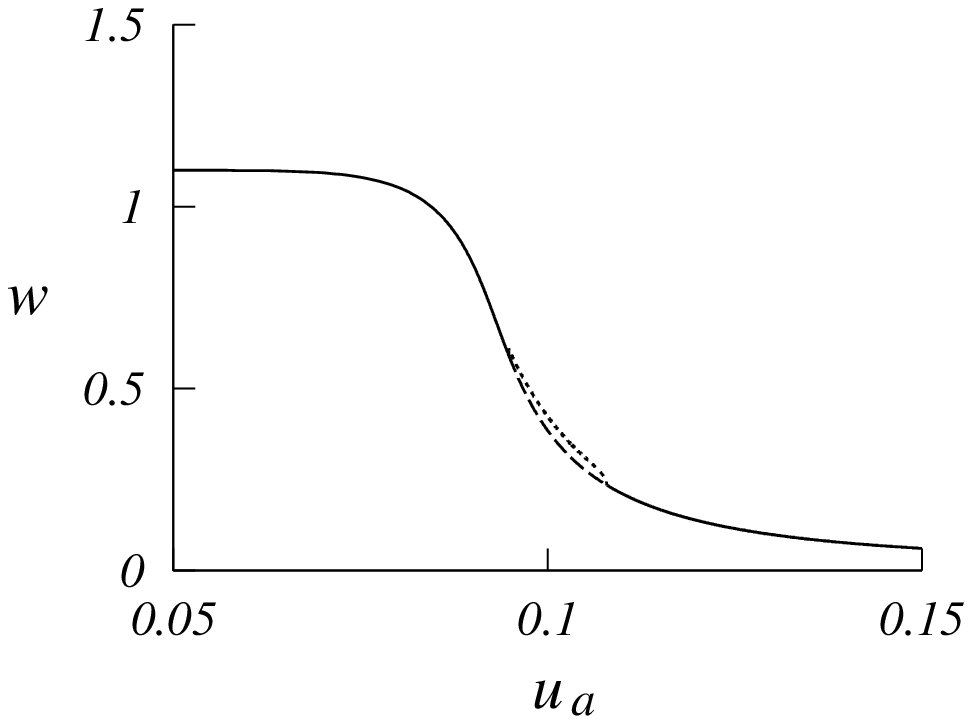}
      \hspace*{-0.5cm}\includegraphics[width=6cm]{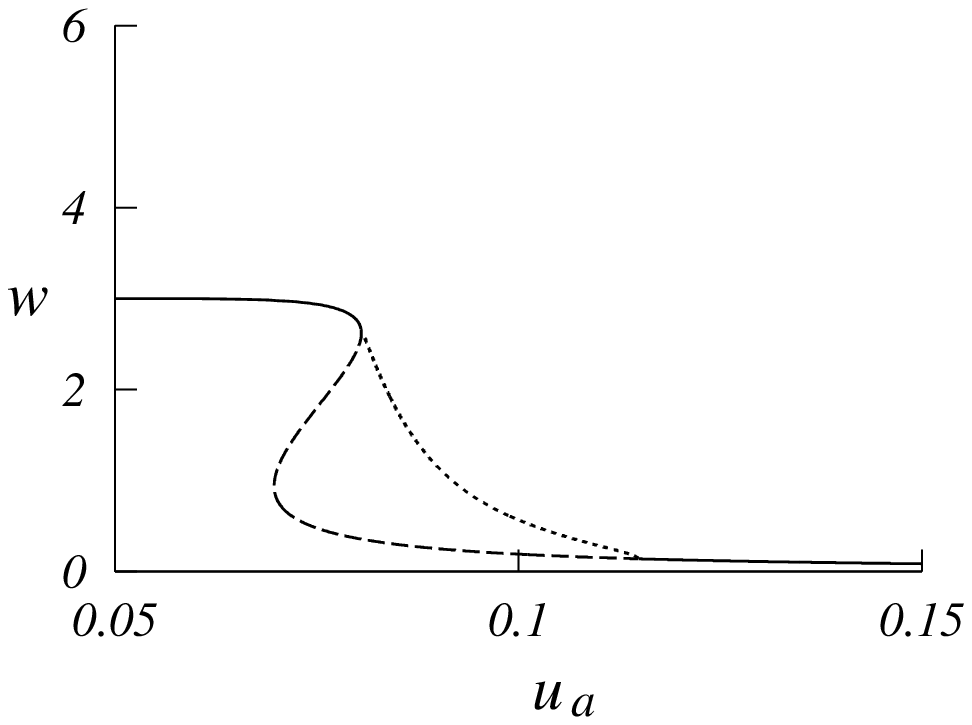}
}
      \hspace*{-1.5cm}
\hbox{\includegraphics[width=6cm]{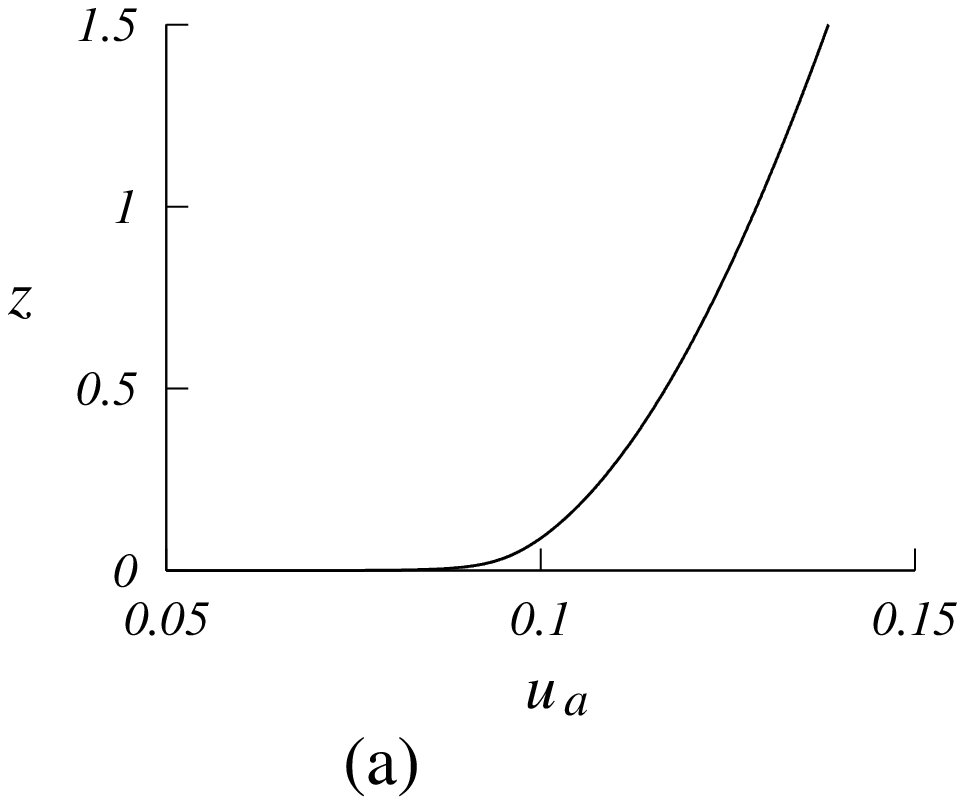}
      \hspace*{-0.5cm}\includegraphics[width=6cm]{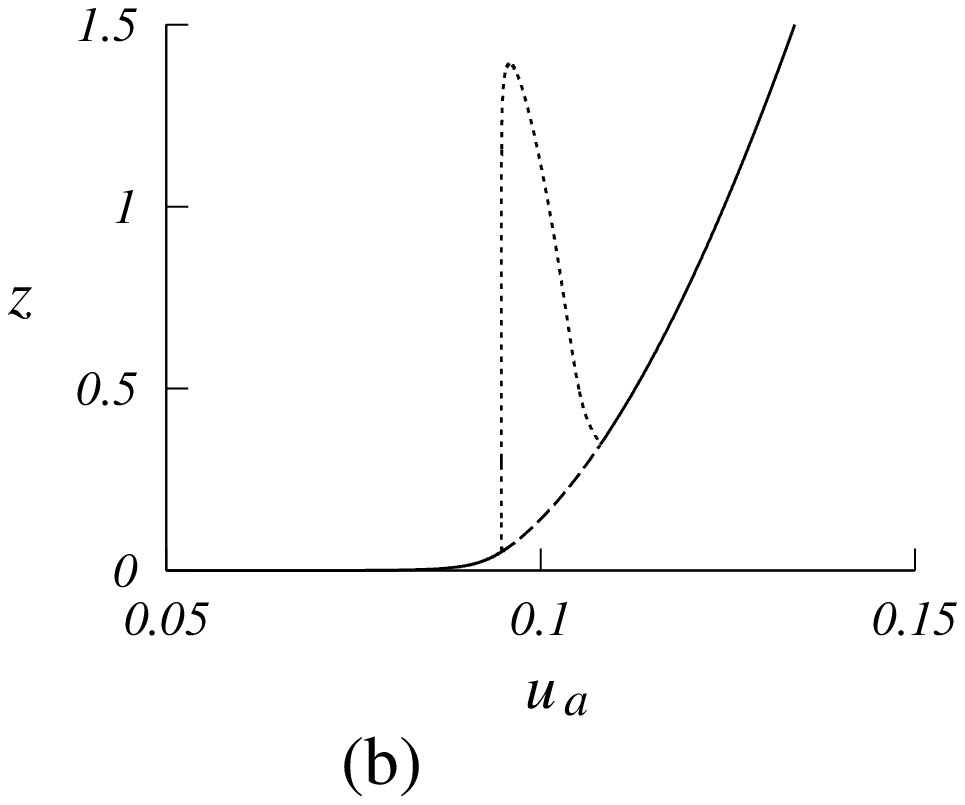}
      \hspace*{-0.5cm}\includegraphics[width=6cm]{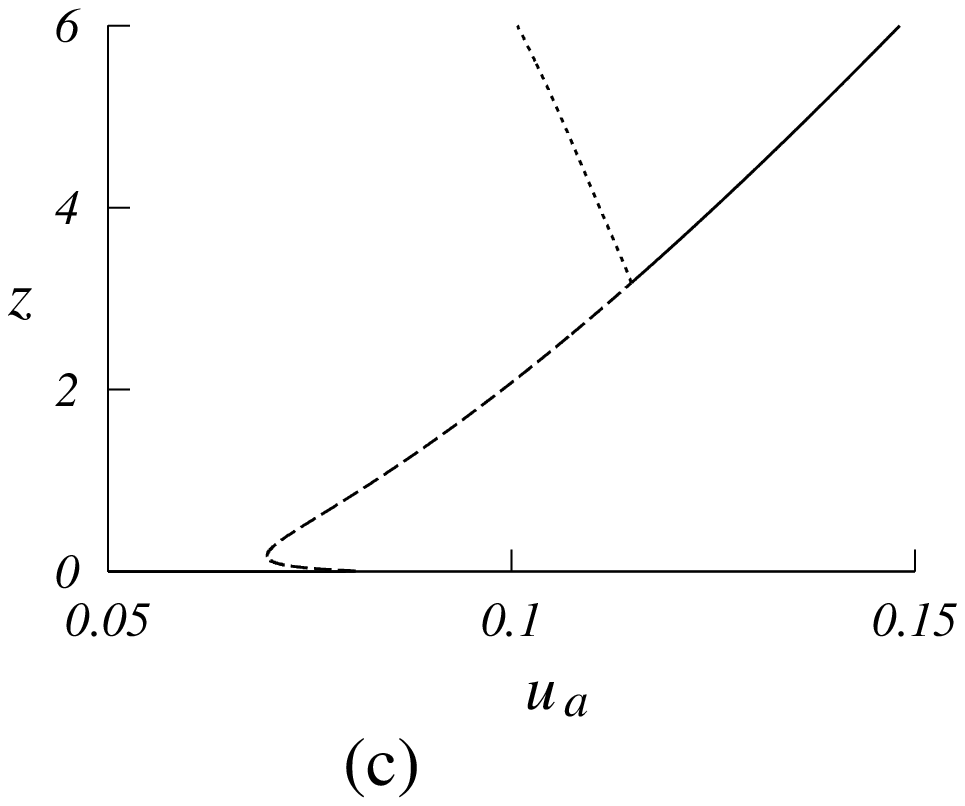}
}\vspace*{1.5cm}            
\caption[Steady states of Eqs. \ref{dwdt}--\ref{dudt} as a
function of $u_a$, the ambient temperature, for 3 values of $r$, the
supply rate of water. (a) $r=0.9\times 10^{-4}$, (b) $r=1.1\times 10^{-4}$,
(c) $r=3.0\times 10^{-4}$. Other parameters: 
$\mu=1.3$, $\alpha=0.3$, $f=0.0001$, $\ell=0.004$, $\bar{C}=1$, $\nu=1$.
Solid lines represent stable steady states, dashed lines are unstable 
steady states, and the dotted lines are the amplitude maxima of the limit 
cycles]{\label{bb}Steady states of Eqs. \ref{dwdt}--\ref{dudt} as a
function of $u_a$, the ambient temperature, for 3 values of $r$, the
supply rate of water. (a) $r=0.9\times 10^{-4}$, (b) $r=1.1\times 10^{-4}$,
(c) $r=3.0\times 10^{-4}$. Other parameters: 
$\mu=1.3$, $\alpha=0.3$, $f=0.0001$, $\ell=0.004$, $\bar{C}=1$, $\nu=1$.
Solid lines represent stable steady states, dashed lines are unstable 
steady states, and the dotted lines are the amplitude maxima of the limit 
cycles.}
\end{figure}

Some steady state solutions of Eqs \ref{dwdt}--\ref{dudt},
computed numerically, 
are shown in Fig \ref{bb}, where 
the dependence on $u_a$ is plotted for three values of $r$. 
Limit cycles, where they exist, have also been computed and plotted as the 
amplitude maxima against $u_a$. ({\it Notes:} (1) Eq. \ref{dzdt} uncouples from 
Eqs \ref{dwdt} and \ref{dudt}. (2) With an assumed activation energy $E_1$ of $\sim$50 kJ/mol
a dimensionless temperature $u$ of $\sim$0.1 corresponds to $\sim$600 K.) 

In (a), where the
supply of reactant is limiting,  the 
temperature rise is monotonic as $u_a$ is increased. When consumption of $w$ 
begins at $u_a\sim$0.08 the rate increases due to thermal feedback, but when
production of $z$ cuts in at a higher temperature the rate levels out again.

In (b) the situation is impressively different, because the steady-state
solutions change stability at two Hopf bifurcations. Here a faster rate of 
supply induces oscillatory behaviour over the
active range of $u_a$. It is noteworthy that, near the onset of the 
oscillations, the concentration of $z$ (the flammable volatiles) and the
temperature can rise dramatically at a relatively low ambient temperature.
The amplitude and temperature range of the oscillations is, of course, also
influenced by the rate of non-reactive heat removal $\ell$ and the rate at
which vapour phase material is removed from the system. However, it is not
difficult to imagine a situation of restricted heat and mass transfer where
this sudden surge in the temperature and concentration of volatiles leads to
spontaneous ignition of the volatiles, whence the system enters a new
regime of flaming combustion. 

In (c) a higher rate of reactant supply induces 
multiple steady states, with a jump occurring at a comparatively low
ambient temperature. Although the upper steady state 
occurs at quite low values of $u$ and $z$, it is unstable, and again $u$ and
$z$ must jump to a catastrophic limit cycle. 
(For clarity the complete branch of limit cycles is not shown in (c), it
terminates in a homoclinic orbit.) The amplitude of the limit cycles 
becomes unrealistically 
high, due to the assumption of an infinite pool of substrate that is built
into this simple model. Nevertheless, the qualitative effect is similar to 
that in (b): the temperature and volatile concentration can surge 
uncontrollably when the competing reaction is initially encouraged. 
\section{Discussion and conclusion}
To a large extent, the development of fire-retarding treatments of and 
strategies for cellulosic materials has been based on the large body
of experimental evidence, some of which was cited above, for the
competitive nature of thermal decomposition. If char is formed at the
{\em expense} of volatile fuel, then intuition suggests that we
should {\em expect} to see a positive
correlation between char yield and fire resistance.   
 Many different additives to cellulosic materials have been found to
 enhance char-formation, e.g., chromated copper arsenate treatments (CCA)
\cite{Helsen:1999}, metal carboxylates \cite{Soares:1998}, sodium hydroxide 
\cite{Chen:1991}, potassium chloride \cite{Jensen:1998}, phosphates 
\cite{Jain:1985}, ammonium salts \cite{Stath:2000}, 
and those studied in \cite{Horrocks:1995} and
\cite{Price:1997}. Some of these are used commercially. 

However, from the point of view of promoting fire safety in real situations, 
the idea that better charring properties equate to better flame resistance 
is too simplistic, because of the nonlinear feedback effects on the competition 
between volatile and char formation described in section 4. 
The results of this study indicate that char-forming treatments
should be used circumspectly, with proper consideration of the particular
situation of potential thermal decomposition --- in particular, whether heat
and volatiles are removed efficiently enough to prevent the thermal feedback
that promotes volatilization.  Even without taking into account
the {\em detailed} role of the hydrolysis chemistry described in section 
\ref{chemistry}, it
is clear from thermokinetic considerations alone that there is a delicate
balance between the amount of char and the degree of fire resistance,
because the temperature --- and therefore the rate of volatile fuel
formation --- can increase dramatically as the exothermal charring reactions
occur. In \cite{Chen:1991} it was reported that pyrolysis of cellulose
became strongly exothermal when the material was treated with large 
amounts of charring retardant. Another mechanism by which extensive 
charring could become dangerous was described in \cite{Diblasi:1993}:
since char has much higher permeability than the unreacted substrate, flammable
volatiles would flow more rapidly through hot char as it forms. The volatiles
may either undergo secondary exothermal charring in the existing char, 
heating up the system further, or ignite on contact with oxygen.

As pointed out in \cite{Gupta:1999}, many other factors
 such as rate of external heating, heat and mass diffusion, 
secondary chemistry, and the immediate
environment can influence the decomposition process. 
None of these is built in to the simplified, homogeneous model in section
\ref{proxy}, which is a crude and much-reduced simplification of a complex
thermokinetic system. Nevertheless, it does reflect the reality of the effects
of cellulose thermal decomposition (there is nothing fine or subtle
about a catastrophic spontaneous fire in a haystack, either), and it is
consistent with the known important features of the process --- namely, the
competitive pathways, the involvement of water, and the thermokinetics.
It is also not difficult to envisage other situations where dangerous 
self-heating could take place. It is often a statutory requirement that
cellulosic materials in bedding, furnishings, and textiles in nursing homes 
contain fire-retardants. However, a bedding fire started by a cigarette is
exactly the kind of situation we have in mind --- the necessary moisture
would already be present and the heat of charring cannot 
easily escape, thus provoking the volatilization reactions.

This study of a simple model does not pretend to be quantitatively precise.
Rather, its intent is to elicit important qualitative information concerning the 
dynamical behaviour of a cellulosic
system under thermal stress, and highlight some thermokinetic effects that
could be a consideration in the design of targeted fire-retarding strategies for 
cellulose. Further modelling studies will examine the thermokinetics in the presence
of fire-retardants, and simulate cellulose decomposition behaviour 
in an oxidizing environment, looking at critical conditions for ignition of the
volatile products and the question of whether decompostion is coupled to the
flame chemistry as well as thermal feedback.

\renewcommand{\baselinestretch}{1}\normalsize

\paragraph*{\bf Acknowledgements:} This work was carried out under EPSRC Grant GR/L28142:
 Char-Forming Processes for Fire Retardancy in Polymeric Materials. R. B. is 
supported by an Australian Research Council Postdoctoral Fellowship.
\newpage
\renewcommand{\baselinestretch}{1}\normalsize

%\bibliographystyle{unsrt}
%\bibliography{cse}

\newpage
\subsubsection*{Appendix}
\vspace*{0mm}Eqs \ref{dwdt}--\ref{dudt} have the following dimensioned form, 
with the symbols mapped
to quantities and dimensionless groups as defined in the accompanying table. \\[-12mm]

\begin{align}
V\frac{dc_w}{dt}&=-Vk_1e^{-E_1/RT}c_w+R_w-Fc_w\tag{\ref{dwdt}$^\prime$}\\
V\frac{dc_z}{dt}&=Vk_2 e^{-E_2/RT}-Fc_z\tag{\ref{dzdt}$^\prime$}\\
VC_{\text{av}}\frac{dT}{dt}&=V(-\Delta H_1)k_1e^{-E_1/RT}c_w+V(-\Delta H_2)k_2 e^{-E_2/RT} +
L\left(T_a-T\right).\tag{\ref{dudt}$^\prime$}
\end{align}
\renewcommand{\baselinestretch}{0.8}\small

\vspace*{-2mm}\begin{tabular}{lll}
{\bf Symbol} & {\bf Definition} & {\bf Units}\\
$V$ & volume of relevant zone & m$^3$\\
$c_w$ & concentration of water in zone & mol/m$^3$\\
$c_z$ & concentration of volatiles in zone & mol/m$^3$\\
$t$&time&s\\
$T$, $T_a$ &system temperature, ambient temperature& K\\
$k_1$ &pseudo first-order dehydration rate constant& s$^{-1}$\\
$k_2$ &pseudo zeroth-order volatilization rate constant &mol/(m$^3$s)\\
$E_1$ &dehydration reaction activation energy&kJ/mol\\
$E_2$ &volatilization reaction activation energy&kJ/mol\\
$R_w$ & constant rate of supply of water from reaction \ref{r1} &mol/s\\
$F$& rate of outflow of vapour-phase species& m$^3$/s\\
$C_{\text{av}}$& average volumetric specific heat of reacting system& J/(m$^3$K)\\
$\Delta H$ & reaction enthalpy& kJ/mol\\
$L$& combined heat transfer coefficient& J/(s K)\\
$R$& gas constant & J/(mol K)\\
$c_{\mathrm{ref}}$& a reference concentration & mol/m$^3$\\
$w$&$c_w/c_{\mathrm{ref}}$&\\
$z$&$c_z/c_{\mathrm{ref}}$&\\
$u$&$RT/E_1$&\\
$\tau$&$tk_1$&\\
$\nu$&$k_2/(k_1c_{\mathrm{ref}})$&\\
$\mu$&$E_2/E_1$&\\
$\alpha$&$\Delta H_2/\Delta H_1$&\\
$\bar{C}$& $C_{av}E_1/(c_{\mathrm{ref}}R(-\Delta H_1))$&\\
$r$&$R_w/(Vk_1c_{\mathrm{ref}}$&\\
$f$&$F/(Vk_1$)&\\
$\ell$& $LE_1/(Vc_{\mathrm{ref}}k_1R(-\Delta H_1))$&
\end{tabular}
%\newpage
\renewcommand{\baselinestretch}{1}\normalsize

\end{document}